\documentclass[preprint2]{aastex}

\shorttitle{Cold Kuiper Belt}
\shortauthors{Gomes et al.}

\begin{document}


\title{Checking the Compatibility of the Cold Kuiper Belt with a Planetary Instability Migration Model}

\author{Rodney Gomes}
\affil{Observat\'orio Nacional \\
Rua General Jos\'e Cristino 77, CEP 20921-400, Rio de Janeiro, RJ,  Brazil}
\email{rodney@on.br}

\author{David Nesvorn\'y}
\affil{Department of Space Studies, Southwest Research Institute, 1050 Walnut Street, Boulder, CO 80302, USA}

\author{Alessandro Morbidelli}
\affil{ Laboratoire Lagrange, UMR7293, Universit\'e C\^ote d'Azur, CNRS, Observatoire de la C\^ote d'Azur, Boulevard de l'Observatoire, 06304 Nice Cedex 4, France}

\author{Rogerio Deienno}
\affil{Instituto Nacional de Pesquisas Espaciais, Avenida dos Astronautas 1758, CEP 12227-010 São Jos\'e dos Campos, SP, Brazil}

\author{Erica Nogueira}
\affil{Universidade Federal Fluminense, Niteroi, RJ, Brazil}

\begin{abstract}

The origin of the orbital structure of the cold component of the Kuiper belt is still a hot subject of investigation. Several features of the solar system suggest that the giant planets underwent a phase of global dynamical instability, but the actual dynamical evolution of the planets during the instability is still debated. To explain the structure of the cold Kuiper belt, Nesvorny (2015, AJ 150,68) argued for a "soft" instability, during which Neptune never achieved a very eccentric orbit. Here we investigate the possibility of a more violent instability, from an initially more compact fully resonant configuration of 5 giant planets. We show that the orbital structure of the cold Kuiper belt can be reproduced quite well provided that the cold population formed in situ, with an outer edge between $44-45$ au and never had a large mass.

\end{abstract}


\section{Introduction}

The Kuiper belt (KB) was initially supposed to be a relic of the original planetesimal disk that could not develop into planets due to a low mass density at great distances from the Sun. It was thus assumed to consist of icy objects which, in absence of strong planetary perturbations, preserved  orbits with low inclinations and low eccentricities.

In 1992, 1992 QB1, the first Kuiper belt object (KBO) was found (apart from Pluto and Charon) and subsequent discoveries revealed a quite excited disk, contrary to the original expectations. Planetary migration theories attempted to explain this excitation of the KB by means of orbital transport processes that would displace planetesimals from the original planetesimal disk to the Kuiper belt region \citep{malhotra1995,gomes2003}. These transport models included scattering of the planetesimals by close encounters with the giant planets and the final implantation of a small fraction of these objects into the Kuiper belt. This emplacement process was made possible by the association of trapping of planetesimals into mean motion and Kozai resonances with a migrating Neptune and a sebsequent escape from these resonances onto stable and excited orbits.

Although many Kuiper belt objects were found on quite excited orbits, some of them had roughly circular orbits on a plane near the ecliptic (or the invariant solar system plane). It did not take too long to collect evidences that there were in fact two distinct populations in the Kuiper belt region, one composed by objects with neutral colors in excited orbits, and another one composed of objects with redish colors in near circular planar orbits. In addition the size distributions of the two populations are different \citep{fraser-a-2014} and the first one is extended to much larger objects than the second. These populations are now respectively known as (dynamically) hot and cold populations of the Kuiper belt. It is generally agreed that the hot population was implanted from the inner regions of the planetesimal disk \citep{nesvorny2015b,gomes2003,levisonetal2008}. However, the origin of the cold population has been debated for a long time. On the one hand, it could be naturally explained as having a local origin, due to its quite different physical properties from the hot population, and because of its cold orbital distribution, similar to that originally expected for a distant undisturbed belt. 

On the other hand, two puzzling features have challanged the local origin hypothesis. The first is the very low current mass of the CKB estimated at $3 \times 10^{-4} M_{\oplus}$ \citep{fraser-a-2014}. Classical planetesimal accretion models would demand much more mass originally in the Kuiper belt region to produce objects as large as those belonging to the cold population \citep{kenyon-bromley2004}. The second feature is that an originally massive disk of planetesimals extending to distances as far as the Kuiper belt region would have displaced Neptune to the outer border of the disk, well beyond the current Neptume's orbit \citep{gomesetal2004}. 

Due to the above difficulty, several scenarios proposed that the CKB did not form in situ, but was implaneted in its current location from a smaller heliocentric distance by some low-efficiency process, so as to explain its current small total mass. Later, \citep{levisonetal2008}  proposed a scenario of implantation of the CKB during the instability of the giant planets and the high-eccentricity phase of Neptune's orbit. However, these scenarios would never quite succeed. One of the problems was the difficulty to decrease the CKB eccentricities to their real values. Moreover these transport models would also require close encounters of the CKB objects with Neptune. These encounters would have unbound the widest binaries, which instead form a large fraction of the current CKB population \citep{parker-kavelaars2010}. 

A possible solution of the problem of the small CKB mass comes from new planetesimal formation theories \citep{youdin-goodman2005,johansenetal2007,Drazkowska-Dullemond2014} that may allow for the production of large objects with a relatively low total mass. If so, a local origin for the CKB becomes again realistic.

Once a local origin of the CKB population is accepted, one can constrain the planetary evolution scenarios that would be compatible with the orbital characteristics of the CKB. \cite{dmc2012} attempted to constrain Neptune's past orbital evolutions that would have preserved the relative low eccentricities and inclinations of the cold Kuiper belt objects (CKBO's). A particularly puzzling feature of the Kuiper belt is the so-called `kernel', a concentration of orbits
with semimajor axes $a\simeq44$~au, eccentricities $e\sim0.05$, and inclinations $i<5^\circ$ \citep{petitetal2011}.
\cite{nesvorny2015} suggested that the Kuiper belt kernel can be explained if Neptune's otherwise smooth
migration was interrupted by a discontinuous change of Neptune's semimajor axis when Neptune reached $\simeq 28$
au. Before the discontinuity happened, planetesimals located at $\sim$40 au were presumably swept into
Neptune's 2:1 resonance, and were carried with the migrating resonance outwards. The 2:1 resonance was at
$\simeq$44 au when Neptune reached $\simeq28$ au. If Neptune's semimajor axis changed by fraction of au at
this point, perhaps because Neptune was scattered off of another planet, the 2:1 population would have been
released at $\simeq$44 au, and would remain there to this day. \cite{nesvorny2015} showed that the orbital
distribution of bodies produced in this model provides a good match to the orbital properties of the kernel.
If Neptune migration was conveniently slow after the jump, the sweeping 2:1 resonance would deplete the population
of bodies at $\simeq$45-47 au, thus contributing to the paucity of the low-inclination orbits in this region.


Here we investigate a scenario alternative to \cite{nesvorny2015}, in which the planets start from a more compact configuration and undergo a more violent instability. This occurs when Neptune is well inside $28$ au and causes a temporarily large eccentricity phase of Neptune's orbit. In this sense, our scenario for the planet evolution is more similar to that adopted in \cite{levisonetal2008}, but with the difference that the CKB starts in situ. 

Like \citep{nesvorny2015,nesv-2011,nesv-morb-2012}, we assume that the giant planet system was made of five giant planets and the existence of an external planetesimal disk. We suppose that the planetesimal disk extends to $45$ au and we check whether Neptune would migrate to its outer border. \cite{gomesetal2004} showed that Neptune would migrate to the outer border of the disk. However, that work was done in the framework of a smooth migration, with Neptune in nearly circular orbit. It is not expected that the same behavior would occur in the case of a global instability featuring close encounter between planets \citep{tsiganis-a-2005,gomes-a-2005} because the disk becomes quite excited and hence it is less effective in driving planet migration. As a second step, we compare the final orbital distribution in the CKB region with the observed one, assuming different initial disk edges within 45 au.  

This paper develops as follows. In Section 2, we describe the numerical integrations we undertake. In Section 3, we analyze the results as far as the final orbits of the planets are concerned. In Section 4, we present an analisys of the best CKBs produced. In Section 5, we analize the results of Section 4 against several constraints. In Section 6, we compare our results with the constraints on Neptune's orbital evolution discussed in \cite{dmc2012} and, in Section 7, we present our conclusions.

\section{The Numerical Integrations }

We performed $2,000$ numerical integrations of the equations of motion of five planets, two of them with their masses like Jupiter's and Saturn's. The other three planets stand for Uranus, Neptune and the fifth ice planet \citep{nesv-2011, nesv-morb-2012}. These share the same mass of $4.5 \times 10^{-5} M_{\odot}$, which is between Neptune's and Uranus' masses. This choice was made since we cannot know a priori, before the integrations ends, which planet will be the outermost one surviving at the end of the migration process. Pairs of neighbor planets from Jupiter to the outermost one are sequentially at their 3:2, 3:2, 4:3 and 5:4 mean motion resonances. \footnote {\ This configuration was obtained through a numerical integration of the four giant planets induced to migrate by an imposed dissipative force proportional to difference between the planet velocity and another factor times the keplerian velocity, like a stokes drag force. This force keeps the planetary eccentricity low, and should thus roughly mimic the torques from the gas disk}. This configuration is significantly more compact than the one considered in \cite{nesvorny2015} and \cite{deiennoetal2017} and perharps more consistent with giant planet migration in a protoplanetary disk of gas \citep{morbidellietal2007}. The initial orbital elements of the planets are shown in Table \ref{elspl}. A disk of planetesimals was placed just outside the outermost planet. In all cases the inner and outer borders are at $16$ au and $45$ au respectively \footnote{the inner edge of the disk was chosen close enough to the outermost planet in order to rapidly trigger the instability phase. We are not here concerned with a possible delay of the instability that might be required to explain the Late Heavy Bombardment as a cataclysmic event \citep{gomes-a-2005}} and they are all formed by $4,000$ equal mass planetesimals. The semimajor axes are chosen uniformly between the inner and outer border of the disk so as to yield a surface mass density varying as $r^{-1}$. The eccentricities were chosen randomly between $0$ and $0.002$. The inclinations were initially null and the other angles are chosen randomly between $0$ and $360^{\circ}$. We adopt four possible total masses for the disk, performing $500$ integrations for each case. These are $25 M_{\oplus}$, $30 M_{\oplus}$, $35 M_{\oplus}$, and $40 M_{\oplus}$, The integrations are performed with the hybrid mercury integrator \citep{chambers-1999}, with a steplength of $0.5$ year and extended to $100$ My. Some of them, which yielded the best final orbital elements for the planets were further extended to $1$ Gy.

\begin{table*}
\begin{center}
\caption{Initial orbital elements for the planets}
\label{elspl}
\begin{tabular}{lccc}
\\
\tableline\tableline
 Planet & semi-major axis (au) & eccentricity & inclination (deg.)\\
\tableline
Jupiter & 5.78788  &  0.00500 &  0.01996 \\
Saturn &  7.81263  &  0.01418 &  0.09230 \\
Core1 &   10.19900 &  0.03912 &  0.02506 \\
Core2 &   12.40130 &  0.01231 &  0.02457 \\
Core3 &   14.48220 &  0.00908 &  0.04247 \\
\tableline
\end{tabular}
\end{center}
\end{table*}

\begin{figure}
\includegraphics*[clip=true,scale=0.3]{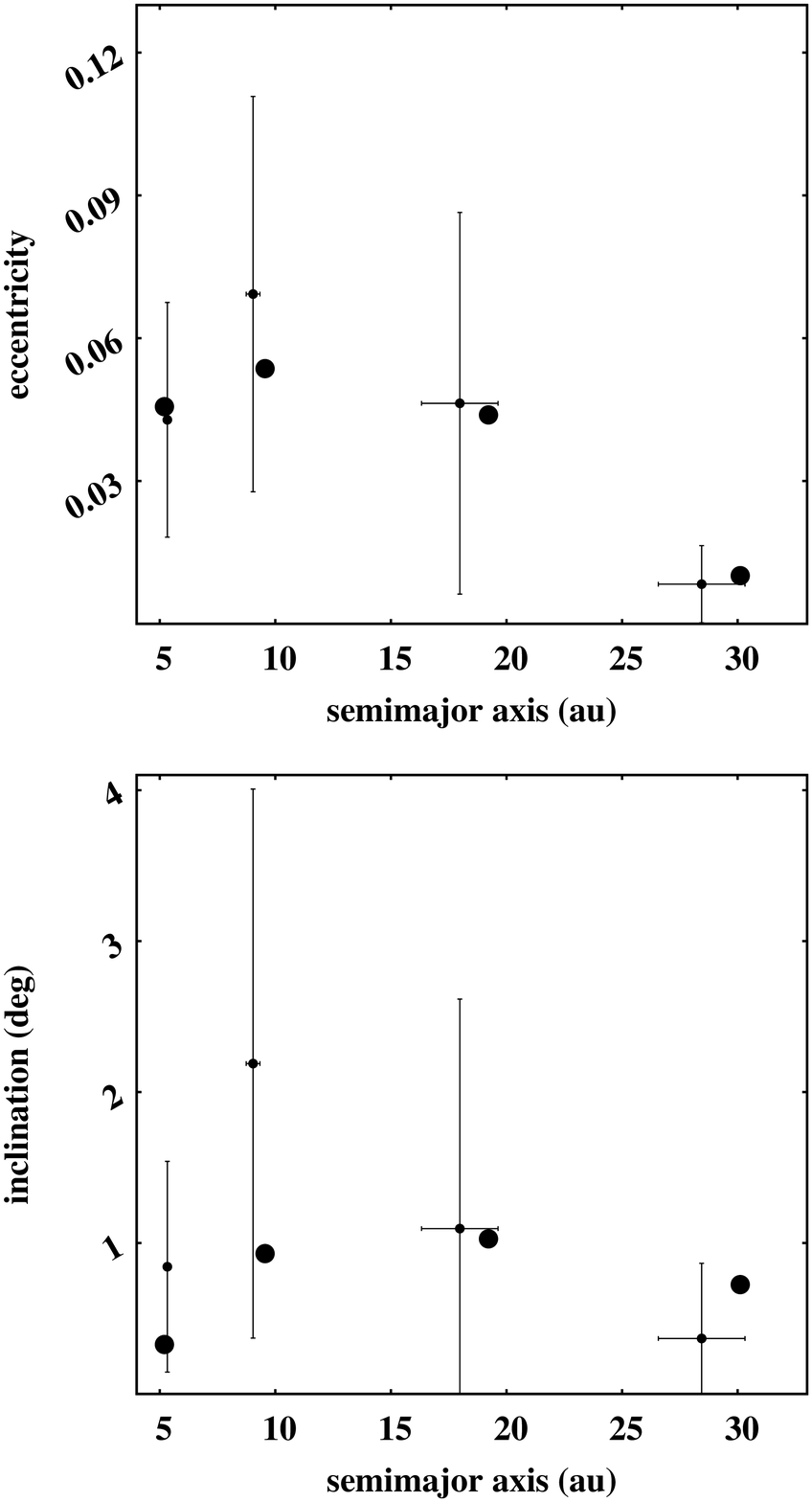}
\caption{Average and standard deviation of semimajor axes, eccentricities and inclinations at $100$ My of the $53$ simulations where the semimajor axes of the planets were in the ranges $5$ au - $5.4$ au, $8.6$ au - $9.9$ au, $15$ au - $21$ au, and $25$ au - $31$ au.. Inclinations are with respect to the invariable plane.}
\label{aeidp-together}
\end{figure}

\section {The final orbits of the planets}

After $100$ My, a fraction of the runs kept exactly four planets. For the disk masses $40$, $35$, $30$ and $25$ $M_{\oplus}$, these numbers are respectively $103$, $96$, $71$ and $67$ in $500$. Restricting the semimajor axes of the four planets at $100$ My in the ranges $5$ au - $5.4$ au, $8.6$ au - $9.9$ au, $15$ au - $21$ au and $25$ au - $31$ au, we are left with $16$, $19$, $9$ and $9$ cases respectively for disk masses $40$, $35$, $30$ and $25$ $M_{\oplus}$, with a total of $53$ cases for all disk masses.

Figure \ref{aeidp-together} shows the average and standard deviation of semimajor axes, eccentricities and inclinations of the planets for the $53$ best cases as above defined. We must allow for some residual migration and circularizarion of the orbits that must take place after the first $100$ My of evolution. Possibly the worst case is for Saturn's inclination. If we consider the median inclination instead of the average one the results are $0.60^{\circ}$, 
$1.57^{\circ}$, $0.50^{\circ}$ and $0.21^{\circ}$, respectively for Jupiter, Saturn, Uranus and Neptune, in better agreement with the real values. This is because the mean is skewed towards large values by a few cases with too high final inclination. Another interesting result is Neptune's semimajor axis at $100$ My around $28$ au in average. As noted above, in an excited disk, Neptune does not necessarily migrate to the outer border of the disk. An example below where the integration was carried out to $4.5$ Gy ratifies this conclusion. 

\begin{figure}
\includegraphics*[clip=true,scale=0.3]{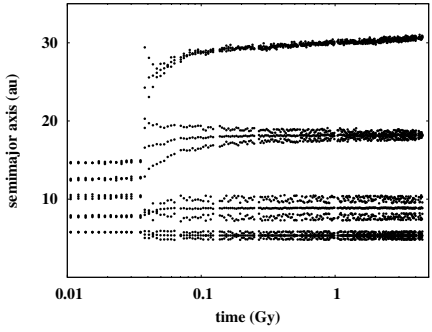}
\caption{Orbital evolution of the semimajor axes, perihelia and aphelia of the planets for one of the simulations which was extended to $4.5$ Gy.}
\label{pls}
\end{figure}

\begin{figure}
\includegraphics*[clip=true,scale=0.3]{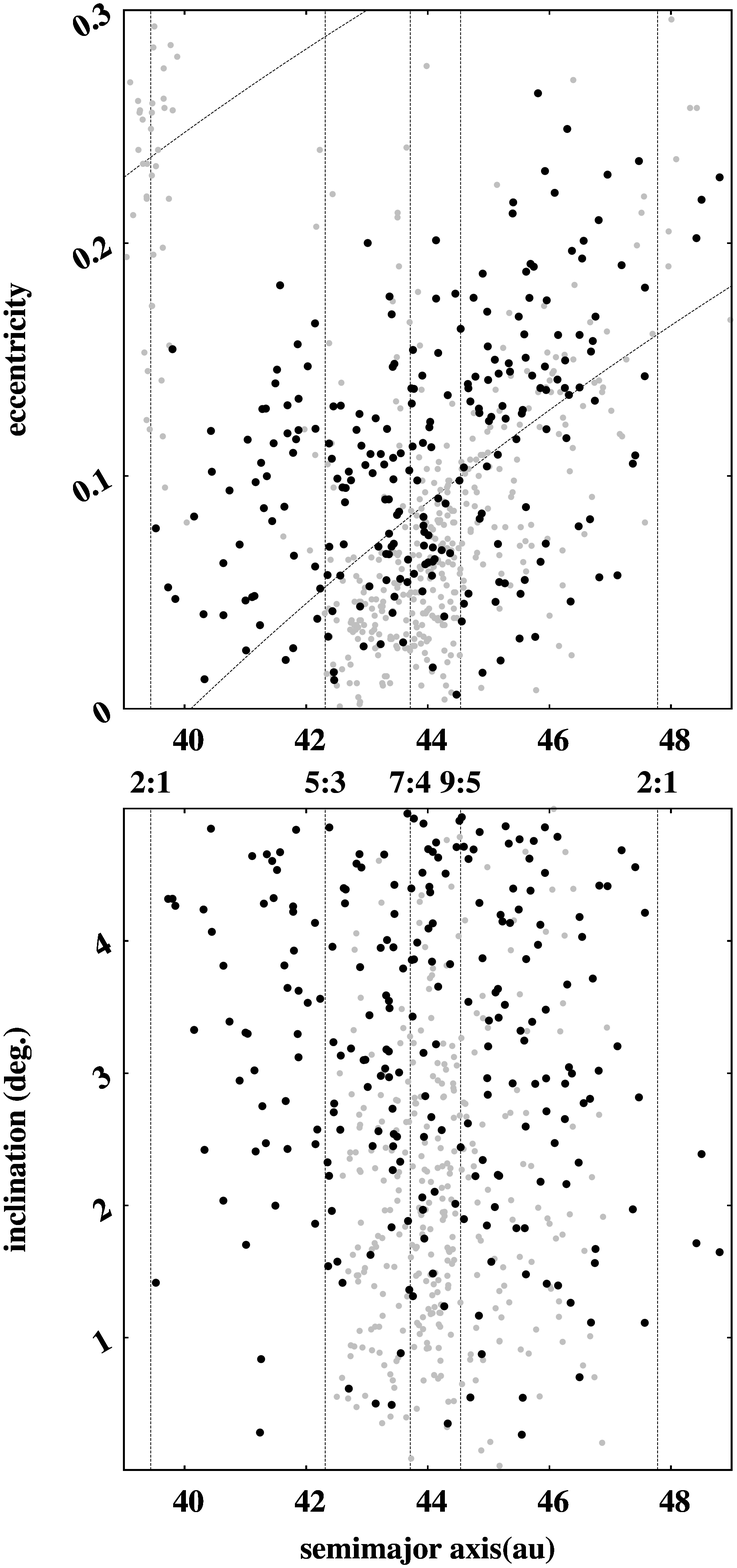}
\caption{Distribution of semimajor axes, eccentricities and inclinations of the CKB at  $4.5$ Gy from one of the simulations (which was extended to $4.5$ Gy) among the $53$ depicted in Fig. \ref{aeidp-together} (black dots) compared with the real CKB (gray dots) }
\label{aeckb0}
\end{figure}

We picked one case of a $30 M_{\oplus}$ disk, where Neptune was at $\sim 29$ au at $100$ My and extended the integration for $4.5$ Gy. At the end Neptune was at $\sim 30.6$ au, thus not far from its actual position.  Figure \ref{pls} shows the evolution of the semimajor axes, perihelia and aphelia of the planets to $4.5$ Gy for that specific case. {\ We note that Neptune is still slowly migrating with a very long migration timescale}. Fig \ref{aeckb0} shows the distribution of semimajor axes, eccentricities and inclinations of the planetesimals at the Kuiper belt region for the same integration at $4.5$ Gy. We just plot the planetesimals with inclinations smaller than $5^{\circ}$ to compare with the CKB \footnote{the real inclinations here and on the rest of the paper are referred to the invariant plane, so as to coherently compare with the simulations}. We notice that the final distribution of the orbits of the planetesimals is not too different from that observed for the real CKB. An important difference is a great number of planetesimals for semimajor axes between $39.5$ and $42$ au. This has no counterpart in the real KB population since the $\nu_8$ resonance is effective in cleaning this region of low inclination objects. But this does not happen in the simulation because the position of the secular resonance depends sensitively on the right positions of the planets which are hard to reproduce exactly in simulations, even though Neptune stopped near its current position. In the next Section we describe the procedure we used to circumvent this problem.

\section {The Cold Kuiper Belt}

We devised the following scheme to determine the cold Kuiper belt (CKB) orbital distribution from the integrations. First we chose the best cases among the $53$ mentioned in the previous section. These cases are those that yielded the best distribution of semimajor axes, eccentricities and inclinations at $100$ My. We were thus left with $12$ cases that showed a reasonable agreement of semimajor axes, eccentricities and inclinations when compared with the real CKB objects \footnote{these are decided by plain visual inspection}. We then continued each of these $12$ integrations until $1$ Gy, in some of the cases setting a null mass for all planetesimals that had started beyond $35$ au in the original integration \footnote{Although Neptune does not usually migrate to the outer border of the disk, it can however migrate too far like to $35$ au. This procedure was taken in order to standardize all runs so as to keep Neptune not too far from $30$ au at the end of the simulations}. After that, new integrations were performed until $4.5$ Gy, restarting from where they stopped at $1$ Gy but changing the planetary orbital elements to the present ones, referred to the invariant plane. The initial orbital elements of the planetesimals in these new integrations were the same as those at the end of the previous integrations to $1$ Gy, except for the semimajor axes which were displaced so as to keep the same period ratio with the real Neptune as those at the end of the integrations to $1$ Gy. No change was done with respect to angular orbital elements of the planetesimals, thus mean motion resonances are not preserved after $1$ Gy. Since we are not aiming at comparing resonant and non-resonant objects in the CKB, we understand that this is a valid approach. In these last integrations all planetesimals are considered as massless particles. These integrations are  important so as to sculpt the simulated CKB with the dynamical features related to the current orbital configuration of the planets. In particular, this procedure allows us to account for the current locations of the secular resonances that are quite active just beyond the 3:2 mean motion resonance with Neptune.

\begin{figure}
\includegraphics*[clip=true,scale=0.3]{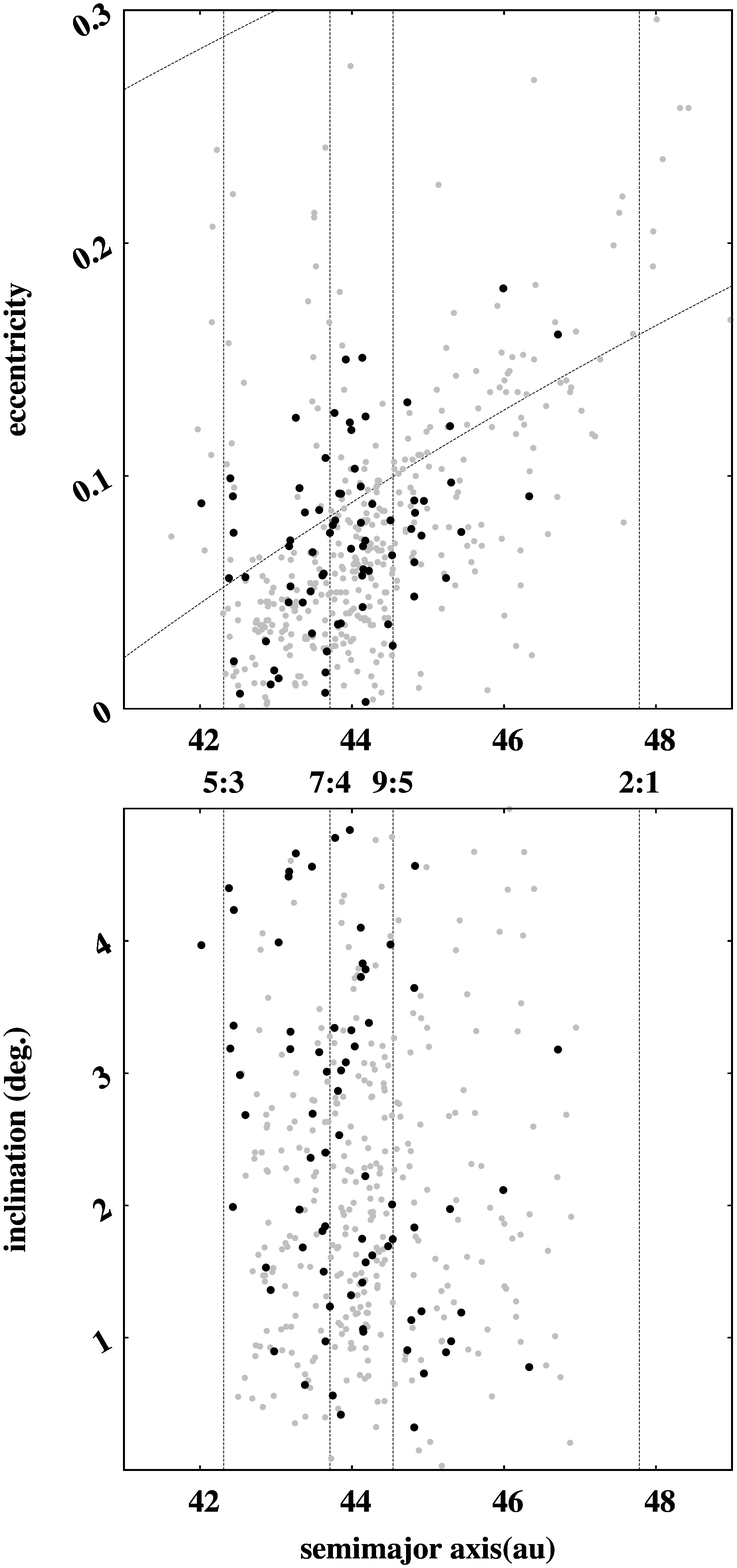}
\caption{Distribution of semimajor axes, eccentricities and inclinations of the CKB from one of the simulations (\#2, see Table \ref{ks}) (black dots) compared with the real CKB (gray dots) }
\label{9-129}
\end{figure}


\begin{figure}
\includegraphics*[clip=true,scale=0.4]{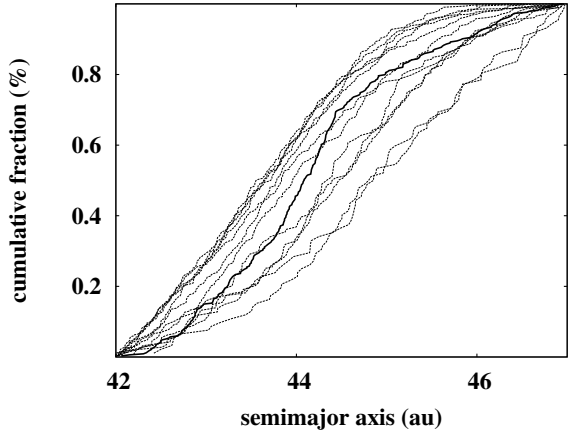}
\caption{cumulative distribution of semimajor axes from all the $12$ simulations, compared with the distribution of semimajor axes from the real CKBO's (thicker line) }
\label{dist-acum}
\end{figure}

\begin{figure}
\includegraphics*[clip=true,scale=0.4]{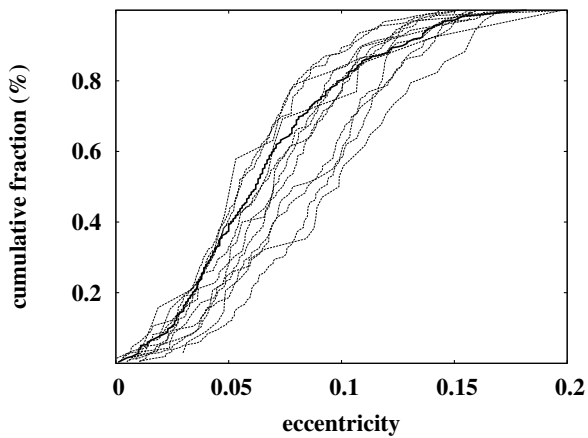}
\caption{cumulative distribution of eccentricities from all the $12$ simulations, compared with the distribution of eccentricities from the real CKBO's (thicker line) }
\label{dist-ecum}
\end{figure}

\begin{figure}
\includegraphics*[clip=true,scale=0.4]{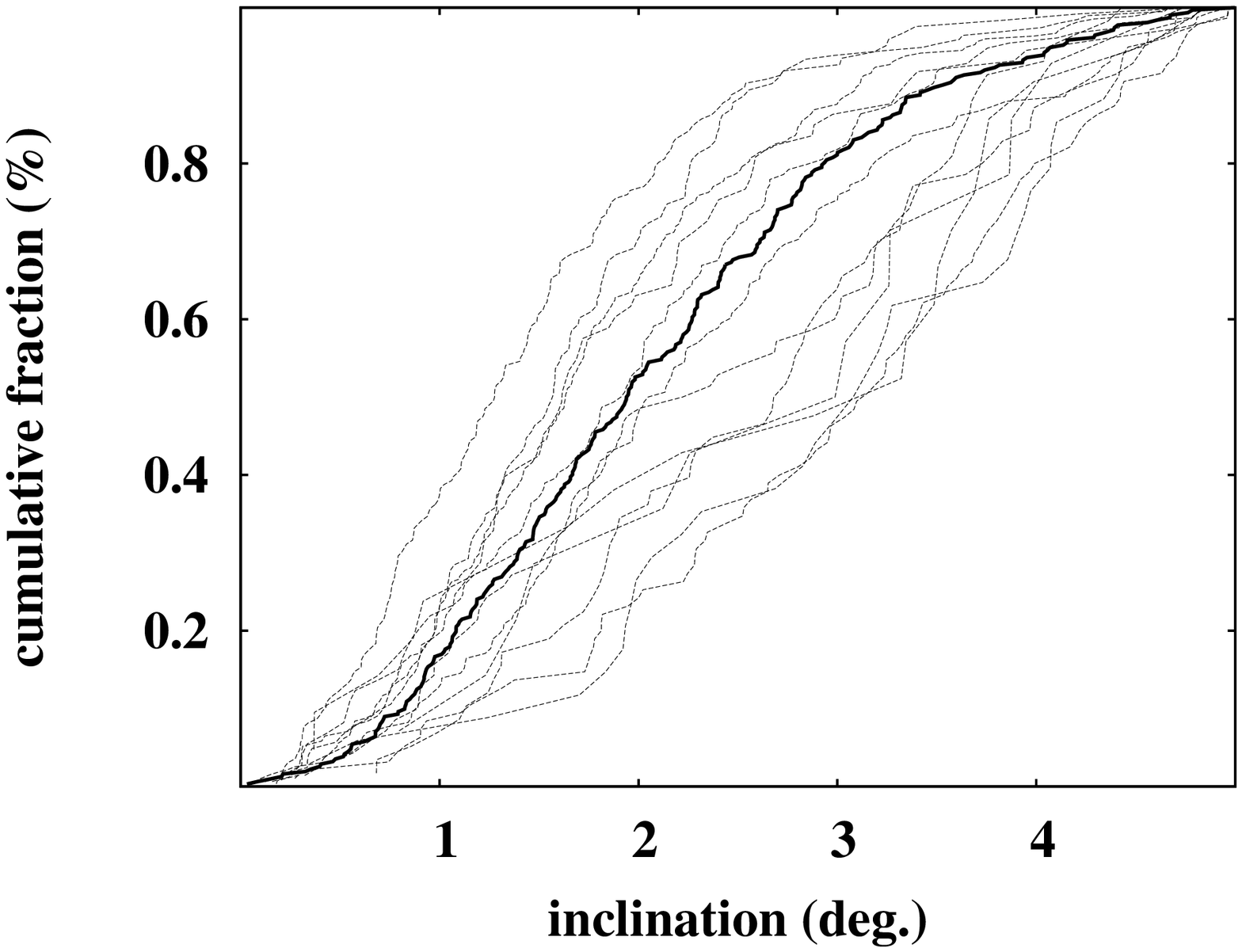}
\caption{cumulative distribution of inclinations from all the $12$ simulations, compared with the distribution of inclinations from the real CKBO's (thicker line)}
\label{dist-i}
\end{figure}

Figure \ref{9-129} shows the distribution of semimajor axes, eccentricities and inclinations for one of the $12$ cases mentioned above at $4.5$ Gy. In these plots we consider orbital inclinations up to $5^{\circ}$, both in the simulated results and the real objects. It is likely that there are real dynamically hot objects with an inclination $i < 5^{\circ}$ mixed up with the cold ones, since these two populations overlap \citep{brown2001}, but we consider this to be a minor effect for the purpose of comparisons. On the other hand, there are also simulated "cold" objects with $i > 5^{\circ}$ which we do not include in the plots. We understand there might be also some real "cold" objects with $i > 5^{\circ}$. Figure \ref{dist-acum} shows the cumulative semimajor axis distribution for the simulated cases compared to the real distribution. The CKB is considered with semimajor axis between $42.5$ and $47$ au, perihelion distance larger than $37.5$ au and inclinations smaller than $5^{\circ}$. Figures \ref{dist-ecum} and \ref{dist-i} show the cumulative distribution in eccentricity and inclination respectively for the simulated cases compared with the real distributions. Although most distributions from the simulations are quite different from the real one, the real distribution is well inside the range of all distributions from the simulations for all three orbital elements.

We performed Kolmogorov-Smirnov (K-S) tests in the semimajor axes, eccentricities and inclinations to compare the simulated CKB populations with the real one. Table \ref{ks} shows the results of these tests. Only cases where more than $50$ objects remain in the CKB are listed, since samples with few components yield less reliable K-S tests. The results are not uniform with respect to the three orbital elements. Possibly the best case is for Integration \#5, although it presents a bad match for semimajor axis. This mismatch might be reduced if the outer border of the disk was changed to some semimajor axis smaller than $45$ au. This is the subject of the next section. 

\begin{table*}
\begin{center}
\caption{Results of the application of a K-S test  comparing the distribution of orbital elements for the simulations and real data. We consider the following range in the elements $42.5 < a < 47$ au, $q > 37.5$ au $I < 5^{\circ}$. We only list the cases with more than $50$ particles. The first column denotes the integration number, the second one the number of CKBO's at the end of each simulation and the following ones the probability that both simulated population and real population comes from the same parent population}
\label{ks}
\begin{tabular}{lcccc}
\\
\tableline\tableline
 Int. & N & $P_a$ & $P_e$ & $P_I$ \\
\tableline
1 & 156 & 0.0000 & 0.0000 & 0.5692 \\
2 & 66 & 0.0874 & 0.3288 & 0.0816 \\
3 & 269 & 0.0001 & 0.0145 & 0.0003 \\
4 & 123 & 0.0385 & 0.1799 & 0.0002 \\
5 & 152 & 0.0017 & 0.8665 & 0.7858 \\
7 & 86 & 0.9921 & 0.0036 & 0.0000 \\
8 & 197 & 0.0010 & 0.0028 & 0.0000 \\
10 & 70 & 0.1827 & 0.2376 & 0.0591 \\
11 & 58 & 0.0000 & 0.0105 & 0.0084 \\
\tableline
\end{tabular}
\end{center}
\end{table*}

\begin{figure}
\includegraphics*[clip=true,scale=0.4]{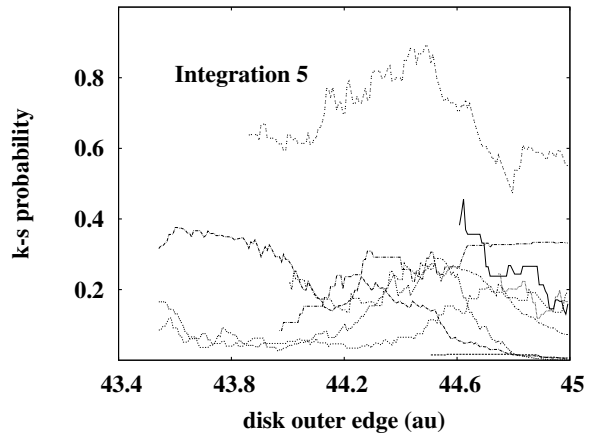}
\caption{The average of the K-S probability in semimajor axis, eccentricity and inclination for all integrations listed in Table \ref{ks} as a function of the disk outer edge. The results of the tests are depicted only if the number of objects in the sample is larger than or equal to $50$.}
\label{paei}
\end{figure}

\begin{figure}
\includegraphics*[clip=true,scale=0.4]{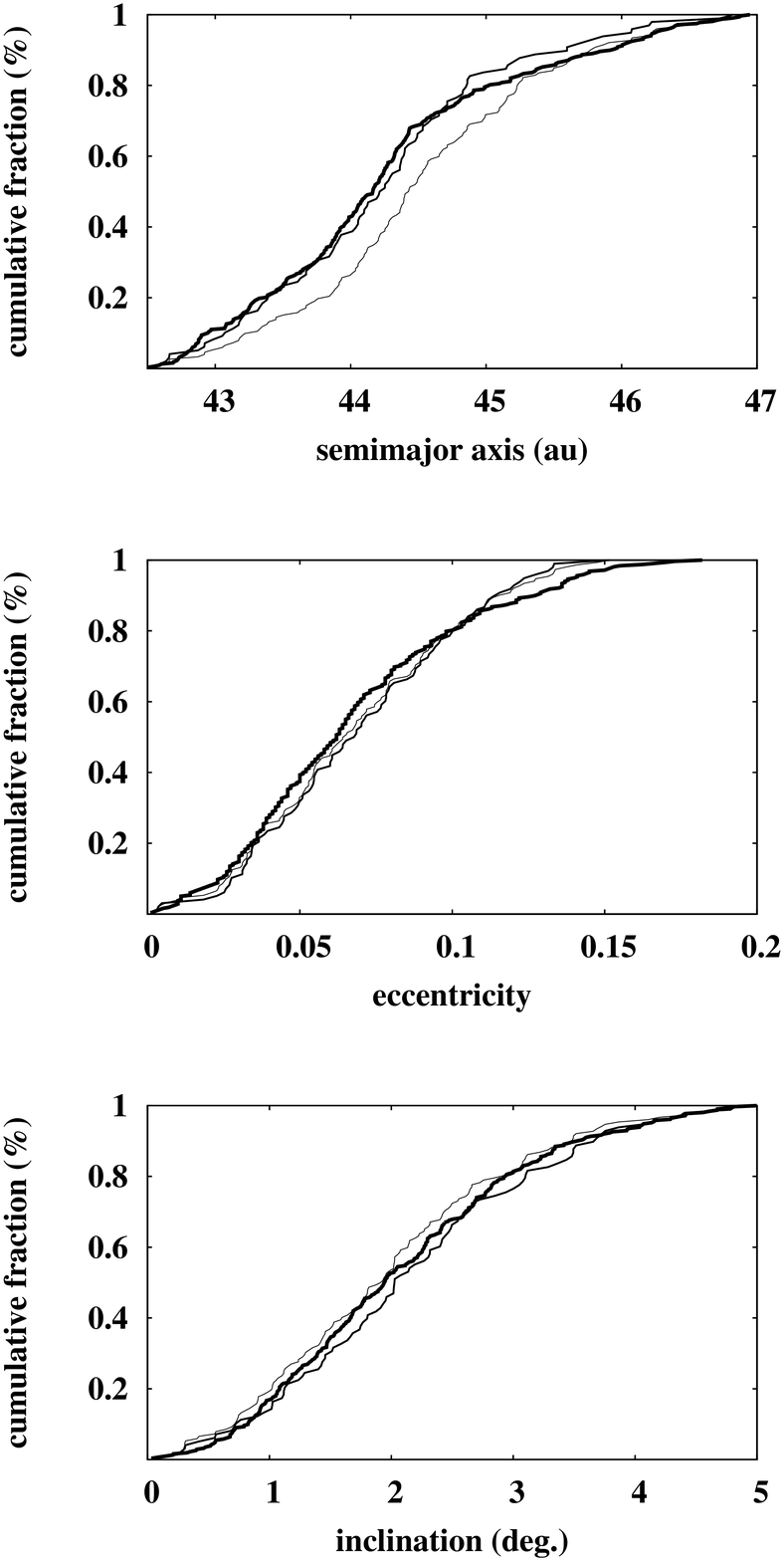}
\caption{cumulative distribution of semimajor axes, eccentricity and inclinations, for the real data (thickest line) and for integration 5 in the case of the outer edge at $45$ au (thinnest line)and at $44.485$ au}
\label{aeicum-3930}
\end{figure}

\subsection{Setting the CKB outer edge as a free parameter}

\begin{figure}
\includegraphics*[clip=true,scale=0.3]{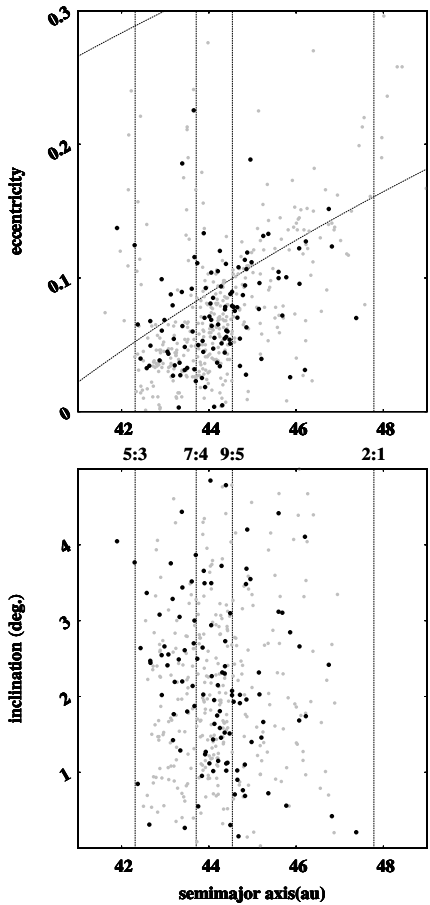}
\caption{distribution of semimajor axes, eccentricities and inclinations for integration \#5, edge at $44.485$ au }
\label{kb2-partsx-9-353-a0-3930}
\end{figure}

The outer edge at $45$ au set as an initial condition for the disk was motivated by the current observed edge around this semimajor axis. Since the real edge is fuzzy, we may conjecture that part of the fuzziness of the edge is due to some dispersion in semimajor axis of the original edge caused by the perturbation of the migrating planets. Since the exact original edges of the disk are not known, we can experiment on the outer edge by discarding from the final data some of the particles that had their initial semimajor axis $a_0 > a_e$ where $a_e$ stands for the outer edge semimajor axis. This procedure was undertaken in order to better compare the CKB determined by the simulations with the real one. The best determined $a_e$ for any specific integration should not be taken as a good determination of the outer edge of the disk. After discarding these particles we test the new final distribution in semimajor axis, eccentricity and inclination of the resultant population comparing with the real distributions through K-S tests. Figure \ref{paei} shows the average of the K-S probability in semimajor axis, eccentricity and inclination for all integrations listed in Table \ref{ks} as a function of the disk outer edge. We just plot the probabilities when there are more than $50$ particles in the sample, since K-S tests turn unreliable for small samples.  We label Integration 5 that yielded the largest values for the K-S average probability for disk edges inside $45$ au. For this case, an outer edge at $\sim 44.5$ au yielded the largest value of the average of the K-S tests in all three orbital elements, as shown in Fig. \ref{paei}. For $a_e=44.485$ au the K-S probabilities for the semimajor axis, eccentricity and inclination are respectively $0.8967$, $0.8282$ and $0.9506$. In Fig. \ref{aeicum-3930} we plot the cumulative distribution of semimajor axes, eccentricity and inclinations, for the real data and for integration 5 in the case of the outer edge at $45$ au and at $44.485$ au. 
Fig. \ref{kb2-partsx-9-353-a0-3930} shows the distribution of semimajor axes, eccentricities and inclinations for integration \#5, after confining the outer edge of the disk at $44.485$ au. {\ Lastly, it is worth mentioning here that in our approach the kernel of the Kuiper belt would be just an outcome of the original outer edge of the disk that was partially scattered but preserved the kernel as a signature of its primordial configuration.}

\subsection{Biasing the simutation data}

The CKB population occupies a fairly compact region in semimajor axes, eccentricities and inclinations. We thus do not expect that the observed CKB diverge too much from the unbiased one. Even so, we here apply a procedure to bias our simulated CKB data. 
We used the CFEPS detection simulator \citep{kavelaars-a-2009} to compare the orbital distributions obtained
in our simulations with observations. CFEPS is one of the largest Kuiper belt surveys with published
characterization \citep{petitetal2011}. The simulator was developed by the CFEPS team to aid the interpretation
of their observations. Given intrinsic orbital and magnitude distributions, the CFEPS simulator returns a
sample of objects that would have been detected by the survey, accounting for flux biases, pointing history,
rate cuts and object leakage \citep{kavelaars-a-2009}. In the present work, we input our model populations
in the simulator to compute the detection statistics.

This is done as follows. The CFEPS simulator takes as an input: (1) the orbital element distribution from
our numerical model, and (2) an assumed absolute magnitude ($H$) distribution. As for (1), the input orbital
distribution was produced by cloning of the final orbits obtained in our numerical model. The cloned orbits
are assigned random values of the perihelion longitude $\varpi$, nodal longitude $\Omega$ and mean longitude
$\lambda$. This procedure should be fine in the region of the cold classicals where mean motion and other
resonances do not play an important role. The magnitude distribution was taken from \cite{fraser-a-2014}.
We varied the parameters of the input magnitude distribution to understand the sensitivity of the results to
various assumptions. We found that small variations of the magnitude distribution within the uncertainties
given in \cite{fraser-a-2014} have essentially no effect.

\begin{figure}
\includegraphics*[clip=true,scale=0.4]{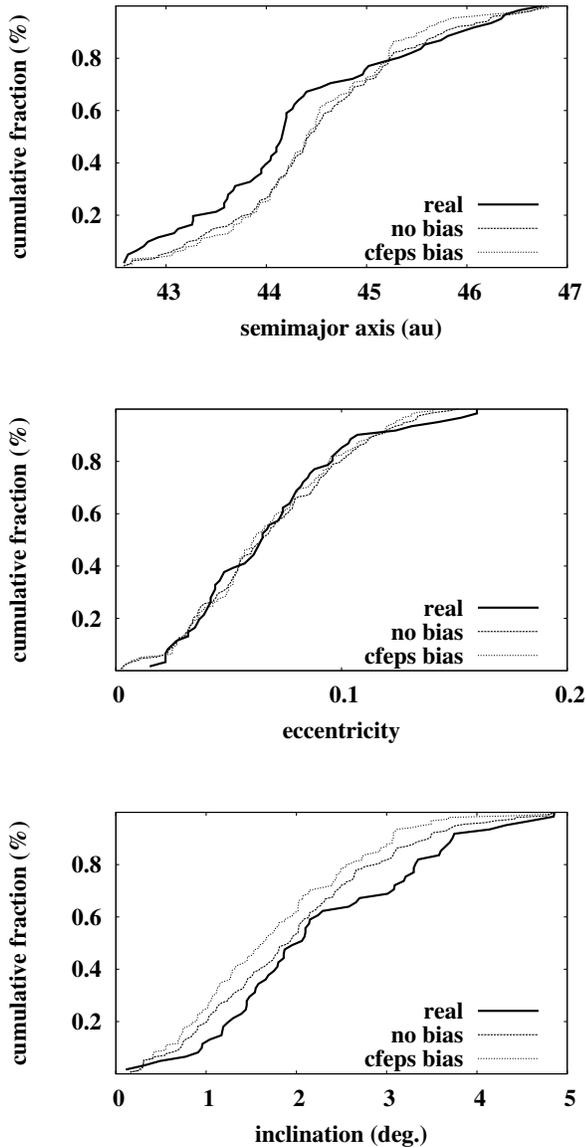}
\caption{Cumulative distribution of semimajor axes, eccentricities and inclinations of the real CKB, compared with an unbiased simulated CKB and the same case after biasing the data, for Integration \#5 with $a_e = 45$ au.}
\label{dist-cfeps}
\end{figure}

After biasing the simulated data we compared the simulated CKB with the {\ CKBOs observed by the CFEPS project} \footnote {orbital elements taken from http://www.cfeps.net/L7Release/CharacterizedList.txt}. Figure \ref{dist-cfeps} shows the cumulative distribution of semimajor axes, eccentricities and inclinations of the real CKB, compared with the unbiased data and the biased one, again for Integration \# 5. We note that the differences of the biased with the unbiased one is very small for semimajor axis and eccentrcitity and a little more significant for the inclination. This example typically represents all the others. The greater difference is in inclination with the biased data showing a distribution skewed to lower inclinations as expected. These tests with the biased data confirm that the comparison with unbiased data is adequate.


\section{Testing the compatibility with other constraints}

\subsection{Keeping the inner solar system intact}

Although our simulations do not place Jupiter and Saturn at the precise distances that would yield their exact present orbital period ratio near $2.5$, we can test how much this ratio varied after the phase of planetary encounters due to residual planetesimal-driven migration. This is an important test since in the real evolution this ratio could not remain below $2.3$ for a long time, otherwise secular resonances would have destabilized the terrestrial planets \citep{brasser-a-2009} and the inner asteroid belt \citep{morbidellietal2010, toliouetal2016}. Among the $12$ cases above studied, in just one of them the orbital period ratio changed by more than 0.2 during the planetesimal driven migration that followed the mutual encounter phase. Because the real planets ended their migration near the 2.5 period ratio, this means that it is likely that the close encounter phase propelled them onto orbits with period ratio larger than 2.3, and then the period ratio increased slowly by less than 0.2. We thus conclude that planetary orbits during migration that yielded good CKB orbits are {\ in principle} compatible with a stable inner solar system. {\ As a caveat, one must note that temporary excitation of Jupiter's and Saturn's eccentricities may also destabilize the inner planets \citep{agnor-lin2012, kaib-chambers2016}.}

\subsection {The Hot Kuiper Belt}

The simulations performed in this paper were not intended to produce the hot population of the KB. Since we have $4,000$ particles in each integration, each particle carries $0.006$ to $0.01 M_{\oplus}$, for disk masses between $25$ and $40 M_{\oplus}$.  Since this latter number is the estimated mass of the hot population \citep{fraser-a-2014}, we conclude that one or at most two particles in each integration should be left in the KB at the end of the integration. This would not be enough for statistical comparisons. We however considered the $12$ good integrations together at $4.5$ Gy and among those planetesimals that started with semimajor axis below $38$ au \footnote{we just demand that the planetesimal is transported from somewhere inside the KB region to the KB to be considered a hot KBO}, we searched for those that ended with semimajor axis between $41$ and $47$ au and $e < 0.35$. We found $18$ that obeyed this criterion, which corresponds to a mass of  $0.012 M_{\oplus}$, thus very close to the estimated mass of the hot population. Their inclinations ranged from $1.4^{\circ}$ to $34.4^{\circ}$, but most of the inclinations ($78$\%) are below $10^{\circ}$. Thus, our hot population is much less excited than the real hot population. We also notice that the range of initial semimajor axes that resulted in the hot KB is $23.3 - 37.6$ au, but most of the planetesimals ($67$\%) come from the range $35 - 38$ au and just $11$\% come from inside $30$ au. We would expect that a larger fraction would come from the inner portion of the original planetesimal disk so as to better explain the physical differences between the hot and cold KBO's. {\ In the future, integrations with a large number of planetesimals in the main disk should be performed so as to better assess the possibility of the production of coherent hot and cold KB populations}



\subsection {The mass of the CKB}

In our scenario the CKB is a local population. The fraction of the initial population surviving in the end is between $0.045$ and $0.65$. Thus, although we have shown that a massive disk of particles extended to $45$ au is compatible with the current orbital location of Neptune, we conclude that the initial mass in the CKB region had to be small. In fact, if the surface density of the CKB had been comparable to that in the inner portion of the disk, about $0.14-2 M_{\oplus}$ would have remained in the CKB today, which is inconsistent with the observations \citep{fraser-a-2014}. This means that the CKB population contained objects up to $300$ km in diameter (about the largest objects currently in the CKB), but carried cumulatively a small total mass. 

This conclusion would be inconsistent with classic model of collisional growth of planetesimals \citep{kenyon-bromley2004}. However, it can be consistent with the new models based on the streaming instability. In fact, \cite{Drazkowska-Dullemond2014} showed that the streaming instability is effective only in regions of the disk where solids could pile up. This could have been inside $30$ or $35$ au. In this case a large fraction of the solid mass is converted into planetesimals, forming a massive planetesimal disk. Instead, in the lower density regions, the streaming instability could operate only near the end of the disk lifetime, when photo-evaporation drastically reduced the gas/solid density ratio \citep{carreraetal2016}. But at the end of the disk lifetime the abundance of solids is not large and hence the overall population produced can have only a small total mass. We believe that our result on the small initial mass of the CKB, which is in agreement with the conclusions of \cite{nesvorny2015}, is a strong argument in favor of the streaming instability model of planetesimal formation. 

\section {Comparing with constraints in \cite{dmc2012}}

We consider Fig. 14 in \cite{dmc2012}. For semi-major axes taking values at $19$, $21$, ...,$29$ au, we evaluate graphically the value of the eccentricity at each of the time curves. This limit eccentricity cannot be surpassed by Neptune's eccentricity for longer than the time indicated by the curve so as to keep the CKBO's forced eccentricities below $0.1$.  
In order to compare with our simulations we considered ranges of $2$ au in semimajor axis from $18-20$ au to $28-30$ au, the center of which are the semimajor axes above chosen. We then compute for how long any of the ice planets keeps its eccentricity above those limiting eccentricities in a specific 2-au range of semimajor axis given by \cite{dmc2012}, by associating the semimajor axis range with its center. We find that in only one of the $12$ cases, an ice planet kept an eccentricity above the limit eccentricity for longer than allowed by \cite{dmc2012} analisys for any of the semimajor axis range. Now if we consider all semimajor axis ranges and sum all the times an ice planet was above the limit eccentricity for each range, we find that in eight cases the total time the ice planet was above the limit eccentricity surpassed the total allowed time. The interpretation of this case can prove harder since the absolute value of the forced eccentricity of a CKBO does not increase continuously as the ice planet semimajor axis changes. In particular, for the case plotted in Fig. \ref{9-129}, an ice planet's eccentricity was above the $10^6$ year-line for a total of $1.14 \times 10^6$ years, as a sum of $10^4$ years in the semimajor axis range $22-24$ au, $2.1 \times 10^5$ years for the range $24-26$ au, $8.4 \times 10^5$ years for the range $26-28$ au and $8 \times 10^4$ years for the range $28-30$ au. Similarly, for the same case, an ice planet's eccentricity was above the $10^{6.25}$ year-line for a total of $1.85 \times 10^6$ years (which is about $10^{6.2672}$ years), as a sum of $2 \times 10^4$ years in the semimajor axis range $18-20$ au, $4 \times 10^4$ years for the range $20-22$ au, $7 \times 10^4$ years for the range $22-24$ au, $4.5 \times 10^5$ years for the range $24-26$ au, $1.19 \times 10^6$ years for the range $26-28$ au, $8 \times 10^4$ years for the range $28-30$ au. Our conclusion is that the restrictions to Neptune's ecentricity proposed in \cite{dmc2012} are a good approximation but do not need to be taken too strictly.

\section{Conclusions}

We performed several simulations of the evolution of initially five planets and a disk of planetesimals located from just outside the furthermost planet to $45$ au. These numerical integrations were initially extended to $100$ My. Several simulations yielded orbits of surviving four planets that were comparable to the current giant planets orbits. These runs were further integrated to $4.5$ Gy to compare the distribution of the further portion of the disk with the present CKB. At the end, we had $12$ runs that reproduced the present CKB fairly well. In some cases Neptune's semimajor axis and eccentricity were dangerously high enough so as to possibly excite too much the CKB, according to \cite{dmc2012}, but they however reproduced a good CKB. The best case showed K-S test probabilities in the distribution of semimajor axis, eccentricity and inclination respectively at $0.0874$, $0.3288$ and $0.7858$. If we set the outer edge of the disk as a free parameter, we have the best case with K-S probabilities at $0.8967$, $0.8282$ and $0.9506$. 

The mass originally at the KB region was depleted from 35\% to 95\% by dynamical erosion. This do not solve the issue of the CKB mass deficiency, but 
alleviate it. The mass initially placed in the planetesimal disk from $42$ au to $45$ au ranged from roughly $2.5 M_{\oplus}$ to $4 M_{\oplus}$ depending on the initial mass of the disk, thus we could get at the end a mass as small as $0.14 M_{\oplus}$ for the CKB, which is still near two orders of magnitude larger than the estimated mass of the CKB \citep{fraser-a-2014}. The orbital distribution we obtain for the CKB at the end is not however highly dependent on the mass at the outer border of the disk, which scarcely interacts gravitationally with the planets. In the future we plan to consider disks with several density distribution profiles so as to turn the migration of the planets and the final mass of the CKB consistent. 

{\ As a last comment we note that we obtain very few cases of formation of a CKB that resembles the real one, possibly just one out of $2,000$ simulations. This suggests that the Solar System is an unlikely outcome from the initial conditions as presented in Table \ref{elspl}. But this may just be pointing to the fact that the Solar System is just a point among a broad range of possible outcomes, any of them quite different from any other} 

\section*{Acknowledgments}

R.D. acknowledges support provided by grants \#2015/18682-6 and \#2014/ 02013-5, S\~ao Paulo Research Foundation (FAPESP) and CAPES. R.S.G. acknowledges his grant no. 307009/2014-9 from CNPq, Conselho Nacional de De-
senvolvimento Cient\'{\i}fico e Tecnol\'ogico, Brazil

\bibliographystyle{apalike}

\bibliography{bib} 

\end{document}